\documentclass[preprint,12pt]{elsarticle}

\usepackage{amssymb}
\usepackage{booktabs} % Allows the use of \toprule, \midrule and \bottomrule in tables
\usepackage{color}
\usepackage{amsfonts,amssymb,amsmath}
\usepackage{amsmath,graphicx}
\usepackage{graphicx,pifont} 
\usepackage{epsfig}
\usepackage{parallel}
\usepackage{esdiff}

\journal{arXiv}

\begin{document}

\begin{frontmatter}



\title{On the microscopic foundation of thermodynamics and kinetics. Current status and prospects.}


\author{A. Yu. Zakharov}

\affiliation{organization={Dept. General and Experimental Physics,\\ Yaroslav-the-Wise Novgorod State University},\\ %Department and Organization
            addressline={41, B. Sanct-Petersburgskaya}, \\
            city={Veliky Novgorod},
            postcode={173003}, 
            %state={},
            country={Russian Federation}}

\begin{abstract}
A comparative analysis of two concepts aimed at microscopic substantiation of thermodynamics and kinetics has been performed. The first concept is based on the idea of microscopic reversibility of the dynamics of a system of particles, while macroscopic irreversibility is of statistical origin. The second concept is based on the idea of the initial microscopic irreversibility of dynamics, one of the mechanisms of which is the relativistic retardation of interactions between particles.

\end{abstract}




\end{frontmatter}


\section{Introduction}
\label{}
By the end of the XIX-th century, phenomenological thermodynamics as ``a branch of mathematical physics founded on the First and Second Laws''~\cite{Guggengeim}, was developed in its perfect form by J. Willard Gibbs~\cite{Gibbs1876, Gibbs1878-1, Gibbs1878-2, Gibbs1906}.

The next stage in the development of thermodynamics is associated with the creation of statistical mechanics (statistical thermodynamics) by Gibbs~\cite{Gibbs1902}, within the framework of which it was possible to build a bridge between classical mechanics and the thermodynamic functions of a system of interacting particles. In addition to classical mechanics, statistical thermodynamics uses an additional hypothesis ~--- the assumption about the possibility of introducing probability measures in the phase space of the system. The main probability measures are three distributions: microcanonical, canonical and grand canonical.

It is significant that when combining deterministic classical mechanics and the concept of probability, the question arises about the mathematical correctness of such a combination. This question arose even earlier within the framework of Boltzmann’s kinetic theory and an exhaustive answer does not yet exist. However, within the framework of Gibbs' theory, this problem was practically not discussed.

The objectives of this work are as follows.
\begin{itemize}
	\item To formulate the main fundamental problems of the physics of many-body and few-body systems, the possibility of an exhaustive solution of which within the framework of statistical mechanics does not seem real. 	
	\item 
	Perform a critical analysis of the assumptions proposed in post-Gibbs studies within the framework of statistical mechanics.
	\item %Обсудить новые точные свободные от вероятности результаты в теории многочастичных и малочастичных систем.
	Discuss new exact probability-free results in the theory of many-body and few-body systems.
	\item 
	Discuss promising directions and open problems in the theory of both many-body and few-body systems beyond the statistical approach.
\end{itemize}

\section{Open problems of thermodynamics}

Let us list some problems of thermodynamics, the solution of which is not possible within the framework of classical statistical mechanics.
\begin{enumerate}
	\item 
	The problem of the existence of thermodynamic equilibrium.
	\item 
	The problem of reconciling deterministic classical mechanics and the concept of probability.
	\item 
	The problem of the existence of probability measures in phase space.
	\item 
	Development of statistical mechanics after Gibbs. %\textbf{Q?!Q}
\end{enumerate}

\subsection{The problem of the existence of thermodynamic equilibrium}

The essence of this problem is as follows:~\cite[p.5]{Uhlenbeck}: ``In its simplest form, one must \textit{explain} in which sense an isolated (that is a conservative) mechanical system consisting of a very large number of molecules approaches thermal equilibrium, in which all \textit{macroscopic} variables have reached steady values. This is sometimes called the \textit{zeroth law} of thermodynamics and it expresses the most typical irreversible behaviour of macroscopic systems familiar from common observation.'' 

The zeroth law of thermodynamics is a postulate of phenomenological thermodynamics and has not found its justification within the framework of statistical mechanics.

\subsection{The problem of reconciling deterministic classical mechanics and the concept of probability}

According to the laws of classical mechanics, the entire evolution of a system of particles is uniquely determined by the equations of dynamics and initial conditions. This completely eliminates probability within the framework of classical mechanics. On the other hand, the observed behavior of many-particle systems is incompatible with reversible and deterministic classical mechanics. Therefore, it is necessary to either ``correct'' classical mechanics, or find a physical reason leading to a ``targeted'' change in the dynamics of a system of interacting particles. Introducing the concept of probability into classical mechanics without elucidating the physical cause of the phenomenon of irreversibility does not solve the problem of microscopic justification of thermodynamics, but complicates it.

\subsection{The problem of the existence of probability measures in phase space}

The fact is that the mathematical concept of probability contains ``hidden'' assumptions. Probability in itself is not a physical quantity and does not allow direct verification~\cite{Newton}. According to the axiomatics of probability theory~\cite{Kolmogorov}, each probabilistic event~$A$ is a subset of the set of elementary events $E$. Random events form a $\sigma$-algebra. The probability of a random event $A$ is a non-negative countably additive measure $\mu\left( A\right) $ that satisfies certain axioms.

Significant restrictions on this axiomatics are imposed by 
{Ulam's theorem}~\cite{Ulam1}: 
``No countably additive measure function $m\left( A\right) $ exists, defined for \textit{all} subsets $A$ of a set $E$ of power $\aleph_{1}$, which vanishes for all subsets consisting of single points and  for which $m\left( E\right) = 1$.''  ~\cite[p.16]{Ulam}.

In this regard, we note that the set of all points of the phase space of a classical system of particles (interacting or non-interacting, it does not matter) is a power set of the continuum. Consequently, there is no single probability measure in the phase space of a particle system. Thus, the status of probability measures, such as the Gibbs distributions, is not impeccable from a mathematical point of view.

As an illustration of Ulam's theorem, we present the well-known Banach-Tarski theorem~\cite{Banach2}, according to which a unit three-dimensional ball can be divided into a finite number of “pieces” and assemble from these pieces two identical balls with a unit radius. The reason for the truth of this paradoxical theorem lies precisely in the existence of Lebesgue immeasurable sets. Therefore, the approach to probability theory based on measure theory leads to a multiplicity of (generally speaking non-equivalent) probabilistic models~\cite{Khrennikov}.

The origin of the paradox is that the axiomatics of probability theory are limited to sets that are Lebesgue measurable. However, Lebesgue immeasurable sets inevitably appear and manifest themselves in probabilistic models related to both mathematics and physics~\cite{Khrennikov1,Khrennikov2}.

Thus, the concept of probability \textbf{ in its present form} is not a completely reliable basis for a microscopic explanation and justification of the laws of phenomenological thermodynamics.

\subsection{Development of statistical mechanics after Gibbs}

Statistical mechanics according to Gibbs establishes a direct connection between the Hamiltonian and the thermodynamic functions of a system of interacting particles. To implement this connection within the framework of the canonical distribution, it is necessary to calculate the configuration integral of the particle system $Z_{\mathrm{conf}}\left(T, V, N, \left\lbrace v\left(\mathbf{r} \right) \right\rbrace  \right) $, which is a function of temperature $T$, volume $V$, number of particles $N$ and functional of the interatomic potential~$v\left(\mathbf{r} \right)$:
\begin{equation}\label{part-func}
	Z_{\mathrm{conf}}\left(T, V, N, \left\lbrace v\left(\mathbf{r} \right) \right\rbrace  \right) = \frac{1}{V^{N} } 
	\idotsint d\mathbf{r}_{1}\cdots d\mathbf{r}_{N}\, e^{-\frac{1}{2k_{B}T} \sum\limits_{j \neq j'} v\left( \mathbf{r}_{j} - \mathbf{r}_{j'} \right) },
\end{equation}
where $k_{\mathrm{B}}$~is the Boltzmann's constant.

It is impossible to perform direct analytical integration in this expression for general interatomic potentials.
Exact solutions to the problem were found only for a limited number of ``unrealistic'' model potentials, among which there is not a single three-dimensional~\cite{Baxter,Sutherland}.
Effective methods for approximate asymptotic calculation in the thermodynamic limit $ N\to\infty,\ V\to\infty,\ N/V = n = \mathrm{const} $ do not currently exist.

There are a limited number of exact analytical results such as the stability criterion for interatomic potentials~\cite{Dobrushin, Ruelle, Fisher}, for which the logarithm of the partition function is an extensive function.

Therefore, a lot of effort is aimed at finding methods for reducing the general problem to simpler \textit{mathematical problems} using additional more or less convincing physical assumptions. The main efforts were concentrated in the following areas.

\begin{enumerate}
	
	\item Bogolyubov hierarchy method~\cite{Bog1} for equilibrium partial distribution functions $F_{s}\left(\mathbf{r}_{1}, 
	\ldots, \mathbf{r}_{s} \right) $ and Bogolyubov, Born, Green, Kirkwood, Yvon~\cite{Bog1, Born, Born1, Kirkwood, Yvon} (BBGKY) hierarchies for non-equilibrium distribution func\-tions $f_{s }\left(\mathbf{r}_{1}, \ldots, \mathbf{r}_{s}; \mathbf{p}_{1},  
	\ldots, \mathbf{p}_{s}; t \right) $.
	Unfortunately, both the equation for the equilibrium partial function $F_{s}\left(\mathbf{r}_{1}, \ldots, \mathbf{r}_{s} \right) $ and the equation for the non-equilibrium partial function $f_{s}\left(\mathbf{r}_{1}, \ldots, \mathbf{r}_{s}; \mathbf{p}_{1}, \ldots, \mathbf{p}_ {s}; t \right) $ are expressed through functions $F_{s+1}\left(\mathbf{r}_{1}, \ldots, \mathbf{r}_{s+1} \right) $ and $f_{s+1}\left(\mathbf{r}_{1}, \ldots, \mathbf{r}_{s+1}; \mathbf{p}_{1}, \ldots, \mathbf{p}_{s+1}; t \right) $ respecti\-vely and therefore not closed.
	To close systems of equations and subsequently solve them, it is necessary to use additional hypotheses such as Bogolyubov’s principle of weakening correlations. However, unlike the self-sufficient theory of Gibbs, the theories of Bogolyubov and BBGKY are not self-sufficient: it is impossible to do without additional hypotheses.

	\item Search for \textit{mathematical} justification of statistical mechanics within the framework of the theory of dynamical systems, including ergodic theory, chaos, attractors, self-organization, etc.~\cite{Kac1,Khinchi1,Chapman,Ruelle2,  Thompson, Balian1, Balian2}. Research in these directions, as a rule, is limited to the framework of non-relativistic classical mechanics. As mechanisms that allow random and irreversible behavior of a system of particles within the framework of classical mechanics, hypotheses such as instability of solutions of dynamic equations with respect to initial conditions~\cite{Zaslavsky}, various versions of ergodic theorems~\cite{Sinai, Weyl, Reed-1980}, baker transformation~\cite{Bricmont, Reichl} etc. are used.

	\item Relativistic thermodynamics and kinetics. After the creation of the theory of relativity, a fundamental possibility (or rather a necessity) arose of bringing thermodynamics and kinetic theory into conformity with the theory of relativity. First of all, the question arises: ``Is there a need to use the concept of probability in relativistic dynamics or does the theory of relativity have its own internal mechanism for the thermodynamic behavior of a system of particles?''
	
	In a non-relativistic theory, a system of interacting particles is completely characterized by its Hamiltonian $H\left(\dots, q_{s}\left( t\right), p_{s}\left( t\right), \ldots \right), $
	depending on \textit{simultaneous} values of coordinates $q_{s}\left( t\right)$ and momenta $p_{s}\left( t\right)$. Therefore, the equations of dynamics of a system of interacting particles are Hamilton's equations.
	However, as is known, a relativistic Hamiltonian of a system of interacting particles, depending on the instantaneous positions of the particles, does not exist~\cite{Currie1, Currie2, Leut}. Therefore, there are only two relativistic models that allow direct implementation of the Gibbs method.
	
	\begin{itemize}
		
		\item Model of an ideal relativistic gas with the corresponding Hamiltonian $ H = \sum\limits_{s=1}^{N} \sqrt{m^{2}c^{4} + p_{s}^{2}c^{ 2}} $. The first works in the relativistic generalization of the kinetic theory of ideal gases were carried out by Planck~\cite{Planck} and \"{J}uttner~\cite{Juttner1,Juttner2}. Synge~\cite{Synge} constructed the relativistic theory of an ideal gas. Further, numerous intensive attempts were made to construct relativistic thermodynamics~\cite{Tolman, Haar, Nakamura}, relativistic statistical mechanics and kinetics~\cite{Chernikov, Groot2, Trump, Cerc1, Hakim, Kuzmenkov1, Liboff, Balescu, Schieve}. In all these works the concept of probability was used by default.
		
		\item Model of contact interaction between particles, in which interaction occurs only at the points of intersection of the world lines of particles~\cite{Dirac, vanDam1, vanDam2}. The scope of applicability of this model is extremely limited to the case of extremely short-range (point) interaction between particles.
		
	\end{itemize}
	
	For constructing the relativistic dynamics of systems of interacting atoms, both of these models are clearly insufficient. As a result, to date it has not been possible to find a unified approach to the construction of relativistic thermodynamics, relativistic statistical mechanics and the relativistic kinetic theory of a system of interacting particles~\cite{Lusanna}.
	
	One of the relativistic effects is the retardation of interactions. In this regard, we note that starting from 1900~\cite{Lamb1} and to date, a number of works have been published in which the dynamics of few-body systems with \textit{retarded interactions between particles}~\cite{Synge1, Driver1, Hsing, Driver2} have been published. It has been established that the retardation of interactions leads to paradoxical (from the point of view of classical mechanics) behavior of systems, including the possibility of an irreversible transition to a state of rest at $t \to \infty$.
	
	Thus, the irremovable retardation of interactions between particles is a real physical mechanism leading to the irreversibility of the dynamics of both few-body and many-body systems. In this regard, to explain and describe the phenomenon of irreversibility there is no need to introduce any additional hypotheses. Note, however, that irreversibility is a necessary, but, generally speaking, not a sufficient condition for explaining and justifying the zeroth law of thermodynamics.
	
	In this regard, it is necessary to develop constructive methods for studying the dynamics of a system of particles with delayed interactions or, more generally, the classical relativistic dynamics of systems of interacting particles.
	
	However, the degree of development of the mathematical apparatus of classical relativistic dynamics in comparison with non-relativistic dynamics is small. Many exact results of classical nonrelativistic dynamics do not take place in relativistic dynamics. In particular, in relativistic theory:
	
	\begin{itemize}
		\item Newton's third law on the equality of action and reaction is violated;
		\item a system of particles with retarded interactions is neither Hamiltonian nor Lagrangian;
		\item there are no integral invariants, including Liouville’s theorem on the conservation of the phase volume of the system;
		\item even the two-body problem does not have an analytical solution.
	\end{itemize}

	Currently, there are two concepts within which it is possible to describe non-instantaneous interactions between particles.
	\begin{itemize}
		\item The theory of direct interparticle electromagnetic interaction, the foundations of which were laid by the works of Tetrode~\cite{Tetrode}, Fokker~\cite{Fokker}, and Wheeler-Feynman~\cite{Wheeler1, Wheeler2}.
		\item Field theory of interactions, according to which the interaction between particles is transmitted through a substance called a field. As a result, a system of interacting particles consists of two subsystems - particles and fields created by these particles. The dynamics of each of these subsystems are described by particle dynamics equations and field evolution equations, respectively. It is important that a system containing even a finite number of particles has infinitely many degrees of freedom due to the field. A well-known example of a field theory of interactions is classical electrodynamics.
	\end{itemize}

\end{enumerate}

\section{Relativistic dynamics of a system of particles within the framework of the theory of direct interparticle interaction}

Due to the retardation of interactions, the relativistic dynamics of a system of interacting particles is described by functional differential equations of retarded type, the general theory of which is still at the initial stage of development~\cite{Berezansky, Xu, Corduneanu}. These equations are essentially infinite-dimensional and have a much richer structure than their counterparts from ordinary differential equations~\cite{Baker}.

\subsection{Theory of oscillations in systems with retarded interactions}

One of the first papers on the dynamics of few-body systems with retarded interactions between particles was carried out by Synge~\cite{Synge1}. In this work, the Kepler problem with retarded interaction was studied and it was shown that in the limit $t \to \infty$ a less massive particle falls on a more massive one and the motion stops.

In the work~\cite{Zakharov2019} the dynamics of a one-dimensional two-particle harmonic oscillator with retarded interaction between particles was studied. It is shown that the characteristic equation of the oscillator has infinitely many complex roots; therefore, non-stationary (both growing and decaying) free oscillations always exist in this system. Thus, the equilibrium state of a two-particle oscillator is stable within the framework of non-relativistic mechanics, but loses stability when the retardation of interactions is taken into account. 

In the work~\cite{Zakharov2021} the dynamics of a one-dimensional chain of atoms with retarded interactions between atoms was studied and it was established that stationary free oscillations in this system are impossible, i.e. the irremovable relativistic retardation of interactions between atoms completely destroys the classical non-relativistic dynamic picture of Born -- von Karman~\cite{Born2,Born3}. In addition, the work~\cite{Zakharov2021} established a microscopic dynamic (i.e., probability-free) mechanism for achieving thermodynamic equilibrium in crystals.

\subsection{Relativistic dynamics of a system of particles with direct interparticle interaction}

The dynamics of a system of interacting particles can be described in terms of \textit{microscopic} (non-averaged) distribution functions defined in 6-dimensional space $\left( \mathbf{r, v}\right) $
\begin{equation}\label{f(r,v,t)}
	f\left(\mathbf{r},\mathbf{v},t\right)  = \sum_{a}\, \delta\left( \mathbf{r}- \mathbf{R}_{a}\left( t\right) \right) \, \delta\left( \mathbf{v}- \mathbf{\dot{R} }_{a}\left( t\right) \right),
\end{equation}    
where $ \mathbf{R}_{a}\left( t\right) $ is radius vector of the $ a $-th particle depending on time~$ t $,
or in terms of covariant microscopic distribution function defined in 8-dimensional space $\left( x, p\right) $
\begin{equation} \label{F(x,p)} 
	{\mathcal F} \left( x,p \right)=\sum _{a}\int\, d\tau _{a}\, \delta^{4}\! \left(  x-x_{a} \left(\tau_{a} \right)\right)  \, \delta^{4} \!  \left( p-p_{a} \left(\tau _{a} \right) \right),
\end{equation}
where $x$ and $p$ are 4-coordinates and 4-momenta, $\tau_{a}$ and $x_{a} \left(\tau_{a} \right)$ are proper time and world line of the $a$-th particle, respectively.

In the works~\cite{Zak-Zub-1, Zak-Zub-2} an exact closed relativistic kinetic equation was obtained for a system of identical classical particles, \textit{direct interaction} between which is carried out through a retarded scalar potential. It is shown that, in the first order in the retardation of interactions, the energy of a system of particles changes over time~$t$ according to the following law
\begin{equation}\label{dE-dt}
	\begin{array}{c}
		{\displaystyle \frac{d}{dt} \left(\sum _{a}\frac{m_{a} v_{a}^{2}\left( t\right) }{2}  +\frac{1}{2} \sum _{a}\sum _{b}U_{ab}   \left(\mathbf{r}_{a}\left( t\right) -\mathbf{r}_{b}\left( t\right) \right)\right) }\\
		{\displaystyle =-\frac{4}{\left(2\pi \right)^{5}c } \int d^{3} \mathbf{k}\left|\mathbf{k}\tilde{\mathbf{j}}\left(\mathbf{k},t\right)\right|^{2}  \int  d^{3} \mathbf{q}\ \frac{\tilde{U} \left( \mathbf{q}\right) } {\left|\mathbf{k - q}\right|^{4} },}
	\end{array}	
\end{equation}
where $\tilde{U} \left( \mathbf{q}\right)$ is Fourier transform of the direct interparticle interaction potential $U\left( \mathbf{r}\right)$, \ $\tilde{\mathbf{j}}\left(\mathbf{k},t\right)$~is Fourier transform of the particle flux density
\begin{equation} \label{j(r,t)-full} 
	\mathbf{j}\left(\mathbf{r},t\right) = \sum _{a}\frac{\mathbf{p}_{a} }{m_{A} }  \delta \left(\mathbf{r-r}_{a} \left(t\right)\right). 
\end{equation} 

It is appropriate to note here that if the function $\tilde{U} \left( \mathbf{q}\right)$ is of constant sign, then the energy of the particle system is a monotonic function of time~$t$.
In statistical mechanics there is a significant limitation on the explicit form of interatomic interactions. It is due to the requirement for the existence of a thermodynamic limit, according to which the logarithm of the partition function of the system must be an extensive function. Interatomic potentials that satisfy this requirement are called stable or non-catastrophic. The criterion for the stability of interatomic potentials was established in the works of Dobrushin, Fisher and Ruelle~\cite{Dobrushin, Ruelle, Fisher}. In terms of the Fourier transformant of potentials for the case of pair interatomic interactions, this criterion has the form~\cite[p.61]{Baus}:
\begin{equation}
	\tilde{U}\left(\mathbf{q} \right)  \ge 0.
\end{equation}
In this case, the energy of the particle system does not increase over time.

This example shows that the relativistic retardation of interactions is a real physical mechanism that can lead to the thermodynamic behavior of the system.

\section{Relativistic dynamics of a particle system within the framework of the field theory of interatomic interactions}

Let us consider a model of a system consisting of neutral particles (atoms), which in the nonrelativistic approximation is characterized by a two-body central scalar interatomic potential of the general form $v\left( r\right) $, which can be represented in the form of a Fourier integral
\begin{equation}\label{Fourier}
	v\left( r\right) = \int\, \dfrac{d \mathbf{k} }{\left(2\pi \right)^{3} }\, \tilde{v}\left(k^{2} \right) \, e^{i \mathbf{k r}},
\end{equation}
where
\begin{equation}\label{Fourier1}
	\tilde{v}\left(k^{2} \right) = \int\, d\mathbf{r}\, v\left( r\right) \, e^{-i \mathbf{k r}}.
\end{equation}

This potential serves as the starting point for the transition from static interatomic potentials to an auxiliary relativistic dynamic field, which is equivalent to interatomic potentials only in the static mode.

The problem is to move from the static potential~$ v\left( r\right) $ to the dynamic auxiliary field $\varphi\left(\mathbf{r}, t \right) $, which satisfies the following requirements:
\begin{enumerate}
	\item the sources of this field are atoms;
	\item in the case of particles at rest, the field $\varphi\left(\mathbf{r}, t \right) $ coincides with $ v\left( r\right) $.
\end{enumerate}

The solution to this problem was obtained in the works~\cite{Zakharov2022, Zakharov2022-2, Zakharov2023} and consists of two stages.
\begin{enumerate}
	\item Based on the static interatomic central potential~$v\left( r\right) $ of general form, which can be represented in the form of a Fourier integral~\eqref{Fourier1}, a partial differential equation for the static auxiliary field is found.
	\item The transition from the equation for a static field to the covariant equation for a dynamic auxiliary field has been completed.
\end{enumerate}

\subsection{Equations of auxiliary fields}

We look for the equation for the static potential~$v\left( r\right) $ created by a particle located at the point $\mathbf{r}=0$ in the form
\begin{equation}\label{f(Delta)}
	f\left( \Delta\right) \left\lbrace  v\left( \mathbf{r}\right)\right\rbrace  = \int \frac{d\mathbf{k}}{\left( 2\pi\right)^{3} }\,  f\left(  -k^{2}\right)  \, \tilde{v}\left( {k^{2}}\right)\, e^{-i \mathbf{k r}} = - 4\pi\, \delta\left( \mathbf{r}\right), 
\end{equation}
where $f\left(\Delta \right) $ is the desired function of the Laplace operator $\Delta$.

Using the Fourier transform, we find
\begin{equation}\label{f(k2)}
	f\left( -k^{2}\right) = -\frac{4\pi}{\tilde{v}\left( {k^{2}}\right)}.
\end{equation}

This relation connects the Fourier transform of the static potential~$ \tilde{v}\left( {k^{2}}\right) $ with the differential equation~\eqref{f(Delta)} describing the corresponding static field.
Thus, the static interatomic potential $v\left( r\right) $, which can be represented as a Fourier integral~\eqref{Fourier}, corresponds to \textit{static field}~$\varphi\left( \mathbf{r}\right) $, which satisfies the linear equation
\begin{equation}\label{varphi(r)}
	\left( \tilde{v}\left( -\Delta\right)\right)^{-1}  \varphi\left( \mathbf{r}\right) = \rho\left( \mathbf{r} \right), 
\end{equation} 
where $\rho\left( \mathbf{r} \right)$ is the density of field sources.

The corresponding homogeneous equation has the form
\begin{equation}\label{varphi-0}
	\left( \tilde{v}\left( -\Delta\right)\right)^{-1}  \varphi\left( \mathbf{r}\right) = 0.
\end{equation}

This equation means that the static field $\varphi\left( \mathbf{r}\right)$ is an eigenfunction of the operator $\left( \tilde{v}\left( -\Delta\right)\right)^{- 1}$, whose eigenvalue is zero. Solutions of the equation~\eqref{varphi-0} describe a {free static fields}.

The first transition from static field equations to dynamic equations {as applied to electromagnetism} was made by L.~Lorenz and Riemann~\cite{Lorenz, Riemann} in 1867, long before the appearance of the theory of relativity. The result is to replace the Laplace operator $\Delta$ in the Laplace and Poisson equations with the D'Alembert operator~$\square$:
\begin{equation}\label{LorRiem}
	\Delta = \dfrac{\partial^{2}}{\partial x^{2}} + \dfrac{\partial^{2}}{\partial y^{2}} + \dfrac{\partial^{2}}{\partial z^{2}}\ \Longrightarrow \	\square = \dfrac{\partial^{2}}{\partial x^{2}} + \dfrac{\partial^{2}}{\partial y^{2}} + \dfrac{\partial^{2}}{\partial z^{2}} - \dfrac{1}{c^{2}}\,\dfrac{\partial^{2}}{\partial t^{2}}.
\end{equation}

Using this Lorentz-Riemann technique, we obtain the equations of the dynamic auxiliary field
\begin{equation}\label{varphi(r,t)}
	\left( \tilde{v}\left( -\square\right)\right)^{-1}  \varphi\left( \mathbf{r}, t\right) = \rho\left( \mathbf{r},t \right), 
\end{equation} 
and 
\begin{equation}\label{varphi-0,t}
	\left( \tilde{v}\left( -\square\right)\right)^{-1}  \varphi\left( \mathbf{r},t\right) = 0.
\end{equation}

Thus, a given central non-relativistic interatomic potential $v\left( r\right) $, which admits a Fourier transform, corresponds to an auxiliary relativistic field $\varphi\left( \mathbf{r}, t\right)$, defined by the equation~\eqref {varphi(r,t)}.

\subsection{Mathematical structure of the equations of free auxiliary fields}

According to the equations~\eqref{varphi-0} and \eqref{varphi-0,t}, the eigenvalue of the operators $\left( \tilde{v}\left( -\Delta\right)\right)^{-1} $ and $\left( \tilde{v}\left( -\square\right)\right)^{-1} $ when acting on a free field is equal to zero. Taking into account the relation~\eqref{f(k2)}, this means that for the corresponding value of $k^{2}$ the function $\left( \tilde{v}\left( k^{2}\right)\right)^{-1} $ becomes zero:
\begin{equation}\label{1/tilde-v}
	\dfrac{1}{ \tilde{v}\left( k^{2}\right)} = 0.
\end{equation}
We will consider this condition as an equation for $k$.

Since the function~$\tilde{v}\left( k^{2}\right)$ for all real values~$k$ is real and has no singularities, then the imaginary parts of all roots of the equation~\eqref{1/tilde-v} must be non-zero:
\begin{equation}\label{k2}
	k_{s} = \alpha_{s} + i \beta_{s} \Rightarrow k_{s}^{2}= \left(\alpha_{s}^{2}-\beta_{s}^{2}\right) + 2 i \alpha_{s}\beta_{s}, \ \beta_{s}\neq 0. 
\end{equation}
In particular, $k_{s}$ can be purely imaginary (for $ \alpha_{s} = 0$), as is the case for the Yukawa potential.

Let us introduce the notation
\begin{equation}\label{k-to-mu}
	\mu_{s}^{2}=-k_{s}^{2}
\end{equation}
and reduce the equation~\eqref{1/tilde-v} to the following form
\begin{equation}\label{1/tilde-v2}
	\dfrac{1}{ \tilde{v}\left( k^{2}\right)} = 	\left(\prod_{s}\left[k^{2} + \mu_{s}^{2} \right]^{\gamma_{s}}   \right)\, F\left( k^{2}\right) = 0, 
\end{equation} 
where $F\left( k^{2}\right) $~ is some function that has no zeros, $\gamma_{s}$~is the multiplicity of the root~$\mu_{s}^{2}$.

\subsection{Factorization of operators }
\begin{enumerate}
	\item 
	For static fields $k^{2}=-\Delta$ and all operators
	\begin{equation}\label{L-s}
		\hat{L_{s}}=\left[ \Delta - \mu_{s}^{2}\right]^{\gamma_{s}}, \quad \hat{L} = \prod_{s}\, \left[ \Delta - \mu_{s}^{2}\right]^{\gamma_{s}}, \quad  F\left(- \Delta \right)
	\end{equation}
	commute with each other. Therefore, the equation~\eqref{varphi-0} is equivalent to the family of equations
	\begin{equation}\label{mu-s}
		\left( \Delta -	\mu_{s}^{2}\right)^{\gamma_{s}} \varphi_{s}\left( \mathbf{r}\right) = 0.
	\end{equation}
	\item For dynamic fields $k^{2}=-\square$ and all operators
	\begin{equation}\label{L-ss}
		\hat{\mathfrak{L}_{s}}=\left[ \square - \mu_{s}^{2}\right]^{\gamma_{s}}, \quad \hat{\mathfrak{L}} = \prod_{s}\, \left[ \square - \mu_{s}^{2}\right]^{\gamma_{s}}, \quad  \mathfrak{F}\left(- \square \right)
	\end{equation}
	also commute with each other. Therefore the equation~\eqref{varphi-0,t} is equivalent to a similar family of equations
	\begin{equation}\label{mu-ss}
		\left( \square -	\mu_{s}^{2}\right)^{\gamma_{s}} \varphi_{s}\left( \mathbf{r}, t\right) = 0.
	\end{equation}
	
\end{enumerate}
For $\gamma_{s}=1$ this equation is the Klein-Gordon equation with the only difference that in the general case the ``mass'' parameter $\mu_{s}$ can be a complex number.

\section{Properties of solutions of the auxiliary fields equations}

The general solution of the auxiliary field equation~\eqref{varphi(r,t)} is the sum of any particular solution of this equation and the general solution of the corresponding homogeneous equation~\eqref{varphi-0,t}. The homogeneous equation~\eqref{varphi-0,t} describes a free field, and a particular solution of the equation~\eqref{varphi(r,t)} contains the field created by atoms.

\subsection{Free field}

\begin{enumerate}
	\item Let us represent the equation of the free dynamic auxiliary field~\eqref{varphi(r,t)} in the following form
	\begin{equation}\label{gen-form}
		\left( \prod_{s'}\, \left[ \square - \mu_{s'}^{2}\right]^{\gamma_{s'}}\right)  \varphi\left( \mathbf{r},t\right) = 0.
	\end{equation} 
	This equation is satisfied by both each of the functions $\varphi_{s}\left( \mathbf{r},t\right)$, and all their linear combinations of the form
	\begin{equation}\label{super-pos}
		\sum_{s}C_{s}\,\varphi_{s}\left( \mathbf{r},t\right).
	\end{equation}

	\item The static auxiliary field $\varphi\left( \mathbf{r}\right)$ also has a similar property. In other words, the central scalar interatomic potential $v\left( r\right) $, which can be represented as a Fourier integral~\eqref{Fourier}, can be represented as a linear combination of {elementary potentials}~--- solutions of the static Klein-Gordon equation or its generalization~\eqref{mu-s}. The parameters $\mu_{s}$ in the static Klein-Gordon equations are in the general case complex.
	In the case when $\mathrm{Im}\, \mu_{s}\neq 0 $, the corresponding elementary static potentials $\varphi_{s}\left( r\right) $ are sinusoidal functions whose amplitudes vary according to Yukawa’s law ~\cite{Zakharov2023}.
	
\end{enumerate}

\subsection{Structure of the auxiliary field of a moving point source}

A particular solution to the inhomogeneous equation~\eqref{varphi(r,t)} can be found using the Green's function method of the Klein-Gordon operator $\hat{\mathfrak{L}}_{s} = \square - \mu_{s} ^{2}$ ~\cite{Zakharov2023}.
The retarded potential of a particle moving according to the law $ \mathbf{r = r}_{a}\left(t \right) $ consists of two parts.

\begin{enumerate}
	\item 
	\begin{equation}\label{rho-1}
		\varphi_{s}^{(1)}\left( \mathbf{r}, t\right) = \frac{1}{4\pi \left( \left|\mathbf{r - r}_{a}\left( \tau\right)  \right|- \frac{\left( \left( \mathbf{r - r}_{a} (\tau)\right) \cdot \dot{\mathbf{r}}_{a} (\tau) \right) }{c} \right) },
	\end{equation}
	where $ \tau $ is a variable related to $ t $ by the relation
	\begin{equation}\label{tau-t0}
		\tau + \frac{\left| \mathbf{r} - \mathbf{r}_{a}\left( \tau\right) \right| }{c} = t.
	\end{equation}
	Potential $ \varphi_{s}^{(1)}\left( \mathbf{r}, t\right) $ at point $\mathbf{r}$ at time $t$ depends on the instantaneous position $\mathbf{ r}_{a}\left( \tau\right) $ and the instantaneous velocity $\dot{\mathbf{r}}_{a} \left( \tau\right) $ of the generating particle at a single moment of time $\tau $, determined by the condition~\eqref{tau-t0}. This potential is an analogue of the retarded Lienard-Wiechert potentials.

	\item \begin{equation}\label{phi2-2}
		\varphi_{s}^{(2)}\left( \mathbf{r}, t\right) = -\frac{\mu_{s}}{4\pi}\int\limits_{0}^{\infty}  \frac{J_{1}\left(\mu_{s} \xi\right)} {\sqrt{\xi^{2}  +  \left|\mathbf{r - r}_{a}\left(\tau\left( \xi, t\right) \right)  \right|^{2} } - \frac{\left( \mathbf{r - r}_{a}\left( \tau\left( \xi, t\right)\right)  \right)\, \dot{\mathbf{r}}_{a}\left( \tau\left( \xi, t\right) \right)  }{c} }\,	d\xi.
	\end{equation}
	
	Potential $ \varphi_{s}^{(2)}\left( \mathbf{r}, t\right) $ at point $\mathbf{r}$ at time $t$ depends on an infinite series of positions $\mathbf {r}_{a}\left( \tau\left(\xi,t \right) \right) $ and an infinite series of instantaneous velocities $\dot{\mathbf{r}}_{a}\left(\tau \left(\xi,t \right)\tau\right) $ of the generating particle at all times determined by the condition
	\begin{equation}\label{tau-k}
		\tau\left( \xi, t\right)   + \frac{1}{c} \sqrt{\xi^{2}  +  \left|\mathbf{r - r}_{a}\left(\tau \left( \xi, t\right) \right)  \right|^{2} } = t
	\end{equation}
	and parameterized by the variable $0\leq \xi < \infty$.
	
	This contribution contains an infinite set of retardations $\tau\left( \xi, t\right)$ between the same points $\mathbf{r}$ and $\mathbf{r}_{a}\left( \tau\left( \xi, t \right) \right) $, depending not only on the distance between these points, but also on the parameter $\xi$.
	
	Retardations $\tau\left( \xi, t\right) \geq \tau_{1}$ correspond to Klein-Gordon waves propagating with velocities from $0$ to~$c$. Note that
	\begin{equation}
		\lim_{\xi \to \infty}\tau\left( \xi, t\right)=\infty,
	\end{equation}
	that is, $\tau\left( \xi, t\right) $ can take arbitrarily large values. This means that the potential $\varphi_{s}^{(2)}\left( \mathbf{r}, t\right)$ is determined not by one point of the particle trajectory, but by all points of the trajectory parameterized by the variable $\xi$.
	
\end{enumerate}

In other words, the observer located at the point $\mathbf{r}$
at time $t$ and using the field $ \varphi_{s}^{(1)}\left( \mathbf{r}, t\right) $ for observation, sees the point source of this field.

The same observer located at point $\mathbf{r}$
at time $t$ and using the field $ \varphi_{s}^{(2)}\left( \mathbf{r}, t\right) $ for observation, instead of a point source of this field sees a source distributed along the trajectory, from infinitely distant past to the point~$\mathbf{r} = \left.\mathbf{r}_{a}\left(t - \tau \left(\xi, t \right) \right)\right|_{ \xi=0} $. This means that the effect of atom $B$ on atom $A$, located at point $\mathbf{r}$ at time $t$, depends on the entire trajectory $\mathbf{r}_{B}\left( \tau\right) $ atom $B$ from the infinitely distant past to the moment of time $\left. \tau\left(\xi, t \right)\right| _{\xi=0} $, determined from the equation~\eqref{tau-k}. This part of the field is almost unpredictable or pseudo-random because it is shaped by the particle's entire prehistory.

\subsection{Qualitative consequences from the relativistic field theory of systems of interacting particles}

The study of qualitative properties of solutions of equations in classical non-relativistic dynamics is greatly facilitated by the existence of exact results such as Liouville’s theorem, integral invariants, conservation laws, etc. Within the framework of relativistic dynamics, these theorems do not hold.

When studying the relativistic field dynamics of a system of interacting particles, the following qualitative properties can be useful.
\begin{enumerate}
	\item 
	The set of degrees of freedom of the system ``particles + field, which ensures the interaction between particles'', is infinite even in the case of a finite number of particles. As a result, a system of interacting particles is no longer a dynamical system with a finite number of degrees of freedom, and specifying initial conditions only for particles is not sufficient for the unique solvability of the Cauchy problem for particles.
	
	\item 
	The dynamics of systems of interacting particles within the field picture depends not only on the equations of motion of particles and the equations of evolution of the auxiliary field, but also on the boundary conditions for the field.
	
	\item 
	An auxiliary scalar field is a superposition of elementary fields, each of which is characterized by its own, generally speaking, complex ``mass'' parameter and satisfies an equation of the Klein-Gordon type or its generalization.
	
	\item 
	The characteristics of the auxiliary scalar field are uniquely determined by the singular points of the Fourier transform $\tilde{v}\left( k\right) $ of the static interatomic potential $v\left( r\right) $ on the complex plane of the wave vector $k$. 
	\item 
	The dynamics of a relativistic system of interacting particles has the property of heredity, since its evolution depends not only on its initial state, but also on its prehistory.
	\item 
	The relativistic effect of retardation of interactions between particles leads to the irreversibility of the dynamics of the relativistic dynamics of both many-body and few-body systems of interacting particles.
\end{enumerate}

\section{Prospects and open problems}

\subsection{Prospects}

Currently, there are two concepts that, in principle, can be used as the basis for the microscopic substantiation of thermodynamics and kinetics.
\begin{enumerate}
	\item 
	The concept of probability, the use of which dates back to Maxwell, Boltzmann, Einstein, Smoluchowski, P. Ehrenfest and T. Ehrenfest, etc. This concept was further developed and implemented in numerous directions. However, the use of the probabilistic concept has two unclear points.
	\begin{itemize}
		\item 	
		The first point is mathematical. It is connected with Ulam's theorem~\cite{Ulam1, Ulam} and Khrennikov's results~\cite{Khrennikov, Khrennikov1, Khrennikov2}.
		
		\item  
		The second point is physical. It is due to the fact that probability is a ``hidden'' non-physical quantity that cannot be measured or verified.
	\end{itemize}

	\item 
	The concept of retardation of interactions between particles (atoms) as a possible cause of irreversibility was expressed by Ritz in a discussion with Einstein~\cite{Ritz}. Interaction delays are a relativistic effect and cannot be doubted. Within the framework of this concept, the following results were obtained, which do not find evidence within the framework of the statistical approach.
	\begin{itemize}
		\item 	
		A dynamic mechanism for achieving a stationary state (that is, macroscopic equilibrium) in the model of a one-dimensional crystal with delayed interactions between atoms has been established.
		
		\item 
		It has been proven that a system of particles with delayed interactions through a stable direct interparticle potential over time irreversibly passes into a state of macroscopic equilibrium~\cite{Zak-Zub-1,Zak-Zub-2}.

		\item  
		It has been proven that the delay of interactions is a general mechanism of the phenomenon of irreversibility for both many-particle and few-particle systems~\cite{Zakharov2019, Zakharov2022, Zakharov2022-2}.

		\item 
		It has been shown that retardations in interactions under certain conditions can lead to synergistic behavior of systems.
	\end{itemize}
	
\end{enumerate}	

\subsection{Open problems}

There are two problems within the relativistic approach.

\begin{enumerate}
	\item 
	The relativistic dynamics of systems of interacting particles is described by functional differential equations of retarded type. In this regard, it is necessary to develop a mathematical apparatus that allows performing both qualitative and quantitative analysis of solutions to such equations.
	
	\item 
	It is not yet known what the connection is between the relativistic dynamics of a system of interacting particles and its thermodynamic functions. This problem must be resolved. 
	Particular attention should be paid to the problem of relativistic dynamics of few-body systems, starting with the two-body problem.

\end{enumerate}

\section{Acknowledgements}

I am grateful to Ya.I. Granovsky, M.A. Zakharov, and V.V. Zubkov for fruitful discussions.





\end{document}